\documentclass{aa}

\usepackage[varg]{txfonts}
\usepackage{epsfig,graphicx,natbib,url,twoopt}
\usepackage[varg]{txfonts}
\usepackage{hyperref}          
\hypersetup{
 colorlinks=true,  
 urlcolor=blue,    
 linkcolor=red,     
 citecolor=blue 
}

\def\kms{\hbox{km$\;$s$^{-1}$}}

\def\Halpha{\mbox{H\hspace{0.1ex}$\alpha$}}
\def\Hbeta{\mbox{H\hspace{0.1ex}$\beta$}}

\def\MgK{\ion{Mg}{ii}~k}

\begin{document}

\title{Ultra-high resolution observations of plasmoid-mediated magnetic reconnection in the deep solar atmosphere}

\author {
Luc H. M. Rouppe van der Voort\inst{1,2}
\and
Michiel van Noort\inst{3}
\and
Jaime de la Cruz Rodr{\'i}guez\inst{4}
}
\authorrunning{L. Rouppe van der Voort et al.}
\titlerunning{Ultra-high resolution observations of magnetic reconnection}

\institute{
  Institute of Theoretical Astrophysics,
  University of Oslo, %
  P.O. Box 1029 Blindern, N-0315 Oslo, Norway
\and
  Rosseland Centre for Solar Physics,
  University of Oslo, %
  P.O. Box 1029 Blindern, N-0315 Oslo, Norway
\and
  Max-Planck-Institut f{\"u}r Sonnensystemforschung, Justus-von-Liebig-Weg 3, 37077 G{\"o}ttingen, Germany
\and
  Institute for Solar Physics, Dept. of Astronomy, Stockholm University, AlbaNova University Centre, 10691, Stockholm, Sweden
}

\date{\today\ - submitted to A\&A Jan 18, 2023 / accepted Feb 22, 2023}


\abstract
{
Magnetic reconnection in the deep solar atmosphere can give rise to enhanced emission in the Balmer hydrogen lines, a phenomenon referred to as Ellerman bombs. 
}
{
To effectively trace magnetic reconnection below the canopy of chromospheric fibrils, we analyzed unique spectroscopic observations of Ellerman bombs in the \Halpha\ line.
}
{
We analyzed a 10 min dataset of a young emerging active region observed with the prototype of the Microlensed Hyperspectral Imager (MiHI) at the Swedish 1-m Solar Telescope (SST). 
The MiHI instrument is an integral field spectrograph that is 
capable of achieving simultaneous ultra-high resolution in the spatial, temporal and spectral domains. 
With the combination of the SST adaptive optics system and 
image restoration techniques, MiHI can deliver diffraction limited observations if the atmospheric seeing conditions allow.
The dataset samples the \Halpha\ line over 4.5~\AA\ with 10~m\AA~pix$^{-1}$, with 0\farcs065~pix$^{-1}$ over a field of view of 8\farcs6~$\times$~7\farcs7, and at a temporal cadence of 1.33~s.
This constitutes a hyperspectral data cube that measures 132~$\times$~118 spatial pixels, 456 spectral pixels, and 455 time steps.
}
{
There were multiple sites with Ellerman bomb activity associated with strong magnetic flux emergence. The Ellerman bomb activity is very dynamic, showing rapid variability and small-scale substructure. We found a number of plasmoid-like blobs with full-width-half-maximum sizes between 0\farcs1 -- 0\farcs4 and moving with apparent velocities between 14 and 77~\kms. Some of these blobs have Ellerman bomb spectral profiles with a single peak at a Doppler offset between 47 and 57~\kms. 
} 
{
Our observations support the idea that fast magnetic reconnection in Ellerman bombs is mediated by the formation of plasmoids. 
These MiHI observations demonstrate that a micro-lens based integral field spectrograph is capable of probing fundamental physical processes in the solar atmosphere. 
}

\keywords{Sun: activity -- Sun: atmosphere -- Sun: magnetic fields -- Magnetic reconnection}

\maketitle

\section{Introduction}
\label{sec:introduction}

Magnetic reconnection is a process that changes magnetic topologies and releases magnetic energy.
It is a physical process that is fundamental for a whole range of dynamical phenomena in magnetized plasmas in environments ranging from laboratories on Earth to the heliosphere and astrophysical bodies. 
In the Sun, magnetic reconnection drives for example solar flares and coronal mass ejections and is invoked in theories that address the heating of the outer atmosphere
\citep[see, e.g.,][]{2014masu.book.....P}. 
In the deeper parts of the solar atmosphere, magnetic reconnection occurs in UV bursts
\citep{2018SSRv..214..120Y} 
and Ellerman bombs \citep[EB,][]{2013JPhCS.440a2007R}. 
In photospheric sites where opposite magnetic polarities are in close proximity, EBs may form and exhibit remarkable enhancements of the spectral wings of the \Halpha\ line
\citep{1917ApJ....46..298E} 
and appear as sub-arcsecond sized brightenings with fine structure and rapid variability in \Halpha\ wing images
\citep[see, e.g.][]{2002ApJ...575..506G, 
2007A&A...473..279P, 
2008ApJ...684..736W, 
2011ApJ...736...71W, 
2013ApJ...774...32V, 
2015ApJ...798...19N, 
2016A&A...592A.100R}. 

While the macroscopic effects of magnetic reconnection are often clearly detectable in the form of for example morphological and structural changes, or in the form of (fast) outflows, details of the microscopic physics remain mostly elusive. 
Of particular interest regarding the mode of reconnection is the formation of plasmoids: small-scale magnetic structures that form when reconnection current sheets break up under the tearing instability into smaller pieces
\citep{1963PhFl....6..459F, 
1986PhFl...29.1520B}. 
This process allows reconnection to happen faster and may explain the observed time scales of reconnection events in the solar atmosphere which are shorter than classical reconnection theory predicts
\citep{2009PhPl...16k2102B, 
2015ApJ...799...79N}. 
Plasmoid formation is in principle a small-scale process that occurs at spatial scales that are beyond the reach of present day instrumentation. However, there have been a number of reports that provide evidence for plasmoids in the solar atmosphere. These are often observations of moving bright blobs at spatial scales near the resolution limit of imaging instruments. 
For example, plasmoid-like blobs have been reported in observations of 
coronal mass ejections \citep{2005ApJ...622.1251L, 
2019SciA....5.7004G}, 
flares \citep{2010ApJ...711.1062N, 
2012ApJ...745L...6T, 
2022NatCo..13..640Y}, 
jets \citep{2012ApJ...759...33S}, 
and rapidly evolving coronal loops \citep{2021NatAs...5...54A} 
Additional evidence for plasmoid formation is based on the evolution and shape of transition region spectral lines spectral associated with UV bursts
\citep{2015ApJ...813...86I, 
2020ApJ...901..148G}. 

The smallest reported plasmoid-like blobs have sizes below 150~km and were observed near EBs associated with UV bursts
\citet{2017ApJ...851L...6R}. 
These EBs occurred in connection with episodes of strong magnetic flux emergence in an active region.
The observed blobs and spectral line profiles were found to be in agreement with numerical simulations of flux emergence where the emerging magnetic field reconnects with the overlying canopy fields. 
A number of recent numerical simulations showed that plasmoids are formed in reconnection events that produce EBs and UV bursts
\citep{2017ApJ...839...22H, 2019A&A...626A..33H, 
2019A&A...628A...8P, 
2021A&A...646A..88N, 
2022ApJ...935L..21N}. 

Ellerman bombs offer a unique window on reconnection in the solar atmosphere: spectral lines in the optical can be employed to resolve the smallest possible scales and get the best possible view on the details of the reconnection process.
This has been recognized as one of the prime science cases 
\citep{2021SoPh..296...70R, 
2019arXiv191208650S} 
for the new 
DKIST \citep{2020SoPh..295..172R} 
and EST \citep{2022A&A...666A..21Q} 
4-m class solar telescopes.
This highly dynamical process poses a formidable observational challenge: structures evolve on a time scale of a few s or less and move with velocities of few tens of \kms\ or more. Typical tunable filtergram instruments that have sufficient spatial and temporal resolution have limited spectral resolution and cannot cover the required spectral range within the dynamical time scale. Typical slit spectrographs that have the required resolution and spectral coverage cannot follow fast moving structures over an extended area. 
Integral field spectrographs can alleviate these problems. 
We analyzed observations from a prototype of a 
microlensed hyperspectral imager \citep[MiHI,][]{2022A&A...668A.149V}, 
a microlens based integral field spectrograph that is capable of acquisition of spatially, spectrally and temporally resolved data over an extended field of view (FOV). 
The dataset samples the \Halpha\ line over a Doppler range of about $\pm$100~\kms\ with 0.5~\kms\ pix$^{-1}$, over a FOV of about 8.6\arcsec$\times$7.7\arcsec\ with 0\farcs065 pix$^{-1}$, during a period of 10~min at a temporal cadence of 1.33~s. 
This unique dataset provides an unprecedented view on the complex dynamics of 
EBs in an active region with vigorous magnetic flux emergence.

\section{Observations}
\label{sec:observations}

\subsection{the MiHI prototype}

The observations were acquired with the Swedish 1-m Solar Telescope 
\citep[SST,][]{2003SPIE.4853..341S} 
where the MiHI prototype was installed as a plug-in for the 
TRIPPEL spectrograph \citep{2011A&A...535A..14K}. 
The MiHI prototype and the methods used to calibrate it and convert the raw data frames into three-dimensional form is described in some detail in a series of three publications: 
\citet[][henceforth VN1]{2022A&A...668A.149V} covering its motivation and design, 
\citet[][VN2]{2022A&A...668A.150V} outlining the modeling and characterization, 
and \citet[][VN3]{2022A&A...668A.151V} detailing the data reduction and image restoration.  
For convenience, we will refer to the prototype simply as ``MiHI'' in the remainder of this paper.

MiHI has a double-sided microlens array 
that samples the image plane and demagnifies each sample such that a 2-dimensional array of point-like sources is passed through the spectrograph. 
Each bright source is then dispersed, generating a spectrum that is separated sufficiently from the spectra from neighbouring sources to avoid significant overlap spatially. 
Spectral overlap between neighbouring spectra is avoided by placing a prefilter in the beam prior to the microlens array 
that limits the wavelength range.

The design of MiHI was optimized for spectropolarimetry in the \ion{Fe}{i} line pair at 6302~\AA\ with a spectral resolution of $R=315\,000$. 
During the 2018 observing campaign, however, spectropolarimetry was also performed in the nearby Na D1 and \Halpha\ lines (see VN3). 
In this paper we concentrate on the \Halpha\ Stokes~I data that were recorded on 24 Aug 2018.

Whenever the seeing conditions allow, the combination of two key factors, the SST adaptive optics system
\citep{2003SPIE.4853..370S, 
2019A&A...626A..55S} 
and image restoration, are able to deliver near diffraction limited spatial resolution in the MiHI data.
The method used for the image restoration is briefly described in VN3, and is based on an image restoration technique developed for slit spectrograph data \citep{2017A&A...608A..76V}, 
which restores the image information by considering that the observed data are the sum of numerous contributions from undegraded object pixels, that are smeared out spatially according to a known point-spread-function (PSF). The resulting linear system connecting the data to the undegraded object pixels was solved using a technique akin to Lucy-Richardson deconvolution. The PSF, required to be known for this technique to work, was determined from the high-cadence phase-diversity images from the context imager using the Multi-Frame Blind Deconvolution 
\citep[MFBD,][]{1994A&AS..107..243L, 
2005SoPh..228..191V} 
wavefront sensing technique.
The context imager was pointed at the MiHI entrance window, consisting of a $2\times2$~mm transparent area of an otherwise black-chromium coated plano-concave lens. 
The reflectivity of both the transparent entrance window and the surrounding coated area were designed to be comparable, so that the context imager receives a comparable amount of the signal from both the MiHI FOV and its surroundings. 
Through the reflection off the MiHI entrance window, the context imager is capable of effectively determining the wavefront over the full MiHI FOV, a prerequisite for the restoration of the spectral data. 
The context imager consisted of two cameras, set up as a phase diversity pair, and operated at a frame rate of 600~Hz and with a 1.5~ms exposure time. 
This was considerably faster than the MiHI camera, which operated at 30~Hz and with a 30~ms exposure time. 
The point spread function (PSF) used for deconvolution of the MiHI data was obtained by averaging over the PSFs of the multiple context exposures with temporal overlap with the longer MiHI exposure, thus satisfying the frozen-seeing condition that is implicitly assumed by the MFBD wavefront sensing method. In this way it is possible to accurately describe the PSF for the 30~ms exposure window of the spectral cameras, while simultaneously maximizing the amount of signal available for the wavefront sensing.
The context imager acquired 20 exposures during each MiHI exposure. 
Since MiHI does not need to scan an extended region like a slit spectrograph, the data can be restored with an arbitrary cadence, in principle up to the frame rate of the camera. The cadence can thus be chosen based on the expected rapid time evolution of the solar atmosphere at the heights sampled by the \Halpha\ line, and was selected to be 1.33~s
(covered by 40 MiHI exposures, 10 per liquid crystal state). 
At such a high cadence, another implicit assumption of the MFBD wavefront sensing method, that the tip and tilt components of the wavefronts vanish when averaged over the period over which the wavefront samples were collected, is no longer satisfied accurately. As a result, the frame of reference of the restored images is still subject to significant residual differential seeing, which manifests itself as a time dependent warping of the image. 

The most obvious way to avoid this problem is to extend the time period over which the context images are collected, and restore the time series at a lower cadence. The spectral data can then be restored using subsets of wavefronts, at the selected cadence of 1.33~s. Unfortunately, the filter used for the context image of MiHI is the same as that used for the spectral camera, and covers only the core and inner wings of \Halpha. Consequently, the image is a relatively even blend of structures from a large range of heights in the solar atmosphere, which is close to ideal from the perspective of providing context for the small instrumental FOV, because it contains an imprint of most structures that are visible in the spectra. From the perspective of image restoration, however, this blend of structures is less than ideal, since their mixing reduces the image contrast, which in turn reduces the reliability of the wavefront sensing. More importantly, however, due to their height in the solar atmosphere, many of these structures evolve on the same short time scale as that expected for the spectra, and restoring these data at a lower cadence does not result in a reliable estimate of the wavefronts. 

To remedy this problem, the technique used by 
\citet{2017A&A...608A..76V} 
to compensate for the scanning motion of the slit in their restoration of scanning slit spectra was employed here as well. To this end, the context data for the entire time series was first reduced at the desired high cadence, but also with significant overlap in time, so that the wavefront coefficients for the same patch of the same frame were determined in the context of different temporal acquisition windows. Since on logical grounds the tip and tilt coefficients of the same patch from the same frame must clearly always have the same value, the difference between the recovered values from two different restorations covering different time periods must be proportional to the displacement between their respective frames of reference. Since many frames were used for each restoration, many such differences can be obtained, the average value of which yields an accurate estimate for the displacements between the patches for each pair of temporally overlapping restorations. 

Integration of these displacements over time yields an accurate estimate of the position of each restored patch, relative to the average position of that patch over the whole time series. This position is subsequently added to each PSF prior to executing the spectral restoration, resulting in a common frame of reference for all restored frames within the time series. With data obtained under good and stable seeing conditions, as was the case for the time series presented in this paper, results of very high stability can be obtained in this way, that need no additional compensation for residual image motion induced by atmospheric seeing or telescope rotation.

An added benefit of using the spectral restoration method is that it can make use of the PSF information to also recover information about object pixels that are located outside the instrumental FOV, giving it the ability to observe ``outside the box''. This ability is a function of the width and variability of the atmospheric PSF in the specific temporal window of the restoration, but it is usually possible to extend the FOV of the instrument by approximately 2 pixels in all directions, dependent somewhat on the required S/N of the restored data. With the 128 $\times$ 114 microlens array elements that were fully sampled by the spectral camera for the present time series, a restored FOV of 132 $\times$ 118~px$^2$ could be obtained.

After restoration, the time series comprises 455 time steps, each represented by a hyperspectral cube of 132~$\times$~118~$\times$~456 spectral pixels (spaxels). With a spatial sampling of 0\farcs065~px$^{-1}$ and a spectral sampling of 10~m\AA~px$^{-1}$, this corresponds to a FOV of 8\farcs58~$\times$~7\farcs67~$\times$~4.5~\AA. The wavelength range runs from 6560.5546 -- 6565.0545~\AA, corresponding to a Doppler offset range of about
$\pm$102~\kms\ around the nominal \Halpha\ line center 
and a Doppler sampling of 0.45~\kms~px$^{-1}$. 
There are a few weak spectral line blends in this spectral range.
Most of these have been identified as telluric H$_2$O lines \citep{1966sst..book.....M}: 
at about $-$78, +32, +57, and +64~\kms. 
Further, there is a weak \ion{Si}{i} line at $-$103~\kms, with an effective Land\'e factor of approximately 1, and a very weak \ion{Co}{i} line at +28~\kms. 

The temporal duration was 10~min and 6~s, starting at 08:03:44 UT on 24 Aug 2018, with a temporal cadence of 1.33~s. 
The seeing quality was measured by the SST adaptive optics wavefront sensor
\citep[see][]{2019A&A...626A..55S}. 
The Fried's parameter $r_0$ for the ground-layer seeing varied between 9 and 34~cm, with the average for the first 6~min above 20~cm. The $r_0$ for a combination of ground- and high-altitude seeing varied between 7 and 10~cm.
The pointing was centered on heliocentric coordinates $(x,y)=(255\arcsec,5\arcsec)$, observing angle $\theta=16\degr$, $\mu=\cos\,\theta=0.96$, targeting the emerging active region AR12720. 
We made much use of 
CRISPEX \citep{2012ApJ...750...22V}, 
a graphical user interface for effective exploration of multi-dimensional datasets and available in SolarSoft. 

The spatial sampling of the context imager was 0\farcs054 px$^{-1}$, and the FOV 49\farcs68 $\times$ 46\farcs82. 
A duplicate of the MiHI pre-filter was used as a filter for the context imager. 
This filter has a bandwidth of about 4~\AA\ and is centered on the \Halpha\ line so that the context images show a complex mix of both photospheric and chromospheric structures. 
The context images were put together in a time series with 915 time steps, at a cadence that is twice as fast as the MiHI spectral data (0.67~s). 
Figure~\ref{fig:context} shows the observed FOV in the context of observations from SDO and IRIS, Fig.~\ref{fig:overview} shows examples of MiHI spectral images in relation to the context image.

\subsection{Reference observations}

\begin{figure*}[!ht]
\centering
\includegraphics[width=\textwidth]{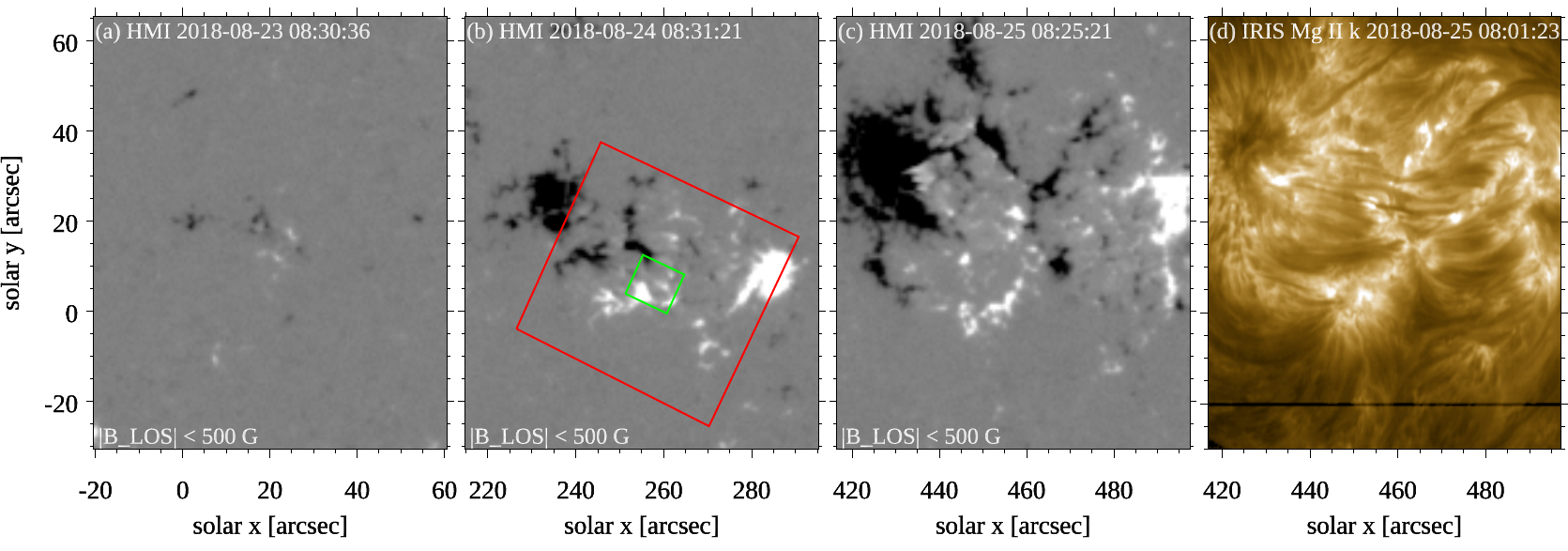} \\
\includegraphics[width=\textwidth]{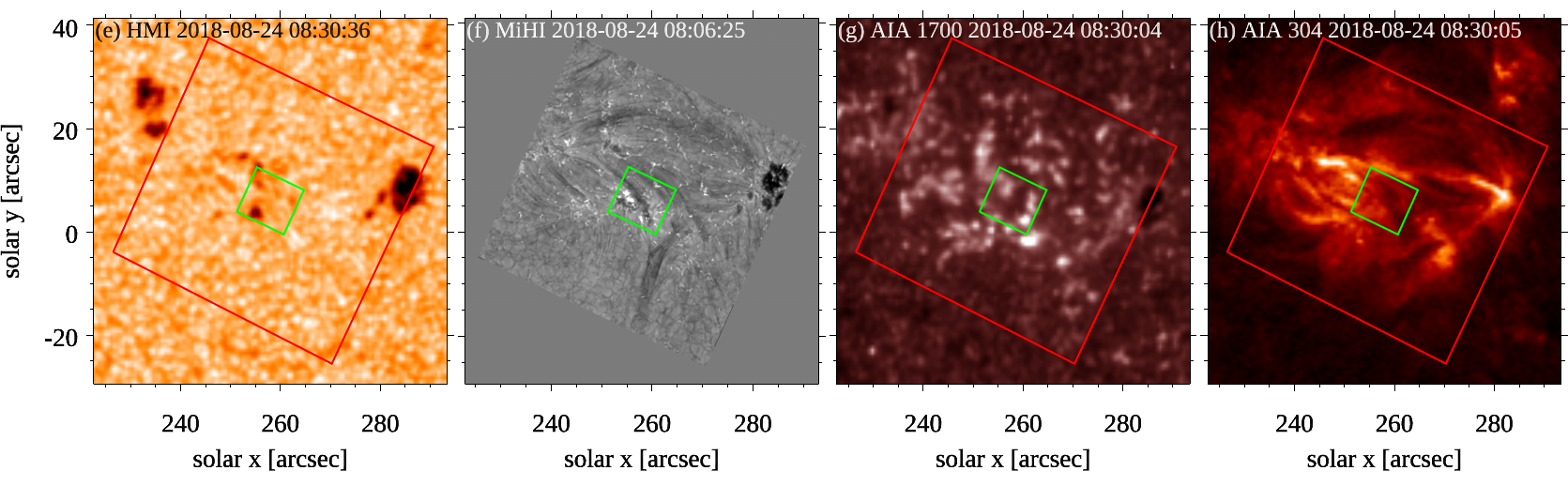}
\caption{
Extended context for the MiHI observations of 24 Aug 2018. Panels (a)--(c) show the emergence of AR12720 in HMI $B_\mathrm{LOS}$ magnetic field maps over a period of 48~h. Panel (b) has the approximate locations of the fields of view of MiHI marked in green and of the context imager in red. Panel (d) shows an IRIS raster map at the nominal wavelength of \MgK\ observed 24~h after the MiHI observations. In the bottom row, the MiHI \Halpha\ wideband context image in panel (f) is accompanied by the closest available SDO images: HMI continuum (e), AIA 1700\AA\ (g), and AIA 304\AA\ (h).
}
\label{fig:context}
\end{figure*}

To put the MiHI observations in an extended context, we studied observations from the 
Solar Dynamics Observatory \citep[SDO,][]{2012SoPh..275....3P}, 
from both the AIA \citep{2012SoPh..275...17L} 
and HMI \citep{2012SoPh..275..207S} 
instruments.
Unfortunately, there were no strictly co-temporal SDO observations since the operations were affected due to an Earth eclipse. 
The earliest available SDO observations of sufficient quality were taken about 20 min after the end of the MiHI observations (see Fig.~\ref{fig:context} panels b, e, g, and h). 
We made extensive use of the JHelioviewer
\citep{2017A&A...606A..10M} 
tool to explore the SDO data. 

We also used 
IRIS \citep{2014SoPh..289.2733D} 
observations of the active region taken the day after the MiHI observations. 
Figure~\ref{fig:context}d shows a \ion{Mg}{ii}~k spectroheliogram constructed from a very large dense 320-step raster (observation identifier \verb|20180825_064806_3620108077|).
This IRIS observation was taken with all four science slit-jaw channels: SJI~1330, 1400, 2796, and 2832~\AA.

We further used a different SST dataset that contains a similar episode of strong flux emergence with anomalously elongated granules as we find in the MiHI observations. 
This episode of flux emergence and associated EBs was part of a 29~min time series of AR12770 observed with the 
CRISP \citep{2008ApJ...689L..69S} 
and CHROMIS \citep{2017psio.confE..85S} 
instruments on 11 Aug 2020 (see Fig.~\ref{fig:textbook}).
From CRISP we used maps of the magnetic field strength along the line of sight (B$_\textrm{LOS}$) derived from Milne-Eddington inversions of the \ion{Fe}{i}~6173\AA\ line.
We used the inversion code developed by \citet{2019A&A...631A.153D}. 
From CHROMIS we used observations in the \Hbeta\ line which was sampled at 27 line positions between $\pm2.1$~\AA, and the associated wideband data (filter width 6.5~\AA, centered at 4846~\AA).
The temporal cadence of the \Hbeta{} data was 17~s. 
The data was processed using the SSTRED reduction pipeline \citep{2015A&A...573A..40D, 
2021A&A...653A..68L}. 
The active region was located at $(x,y)=(321\arcsec,296\arcsec)$ and observed under an observing angle of $\theta$=27\degr\ ($\mu = 0.89$). 

\section{Results}
\label{sec:results}

\subsection{Magnetic flux emergence in AR12720}
The MiHI instrument was targeting the central area of a young emerging active region that was in a state of massive flux emergence. 
As shown in the HMI B$_\mathrm{LOS}$ map in Fig.~\ref{fig:context}a, 24~h earlier, the active region had only just started to emerge to the surface, while 24~h later (Fig.~\ref{fig:context}c), the active region had grown so much that it had received its NOAA AR number: AR12720. 
%
The emergence of the active region over this period was accompanied with enhanced activity in all the AIA channels, for example as bursts of compact brightenings in the AIA 1600 and 1700~\AA\ channels.
The AIA 1700~\AA\ image in Fig.~\ref{fig:context}g shows one such brightening inside and one just outside the area that was covered by MiHI about 25 min earlier. %
These compact UV continuum brightenings are associated with
EBs \citep{2019A&A...626A...4V}. 

The AIA 304~\AA\ channel, dominated by the \ion{He}{ii} resonance line, showed high levels of enhanced emission in and around the emerging region, as well as long dark fibrils that obscure some of the emission from the deeper parts of the atmosphere. 
Some of these dark fibrils are visible in Fig.~\ref{fig:context}h. 
Long dark chromospheric fibrils are clearly visible in the IRIS \ion{Mg}{ii}~k raster map in Fig.~\ref{fig:context}d. 
These fibrils form a chromospheric canopy of pre-existing magnetic fields which the newly emerging magnetic flux encounters from below as it rises through the lower atmosphere. 
These encounters lead to magnetic reconnection and dissipation of currents which is associated with impulsive heating and flaring loops as for example can be seen in the IRIS transition region slit-jaw channels SJI~1330 and 1400~\AA. 

The densest chromospheric fibrils were also visible in the MiHI context images (Fig.~\ref{fig:context}f and large panel in Fig.~\ref{fig:overview}). 
These fibrils had sufficient opacity over the filter passband to effectively block radiation from the deeper atmosphere below. 

From the context SDO and IRIS observations, we conclude that MiHI was targeting the heart of a region in the middle of an extended period of strong magnetic flux emergence. 
Earlier emerged flux had already established an overlying chromospheric canopy of dense fibrils.
New magnetic flux that freshly appeared at the surface just prior to or during the short, 10~min, duration of the MiHI observation was caught in the process of interacting with the pre-existing ambient magnetic field. 
While there are no detailed measurements of the magnetic field and the magnetic field topology is therefore unknown, such a region with strong flux emergence is known to have high occurrence of EBs 
\citep[see, e.g.,][]{2002ApJ...575..506G}. 

\subsection{Ellerman bombs}

\begin{figure*}[!ht]
\sidecaption
\includegraphics[width=12cm]{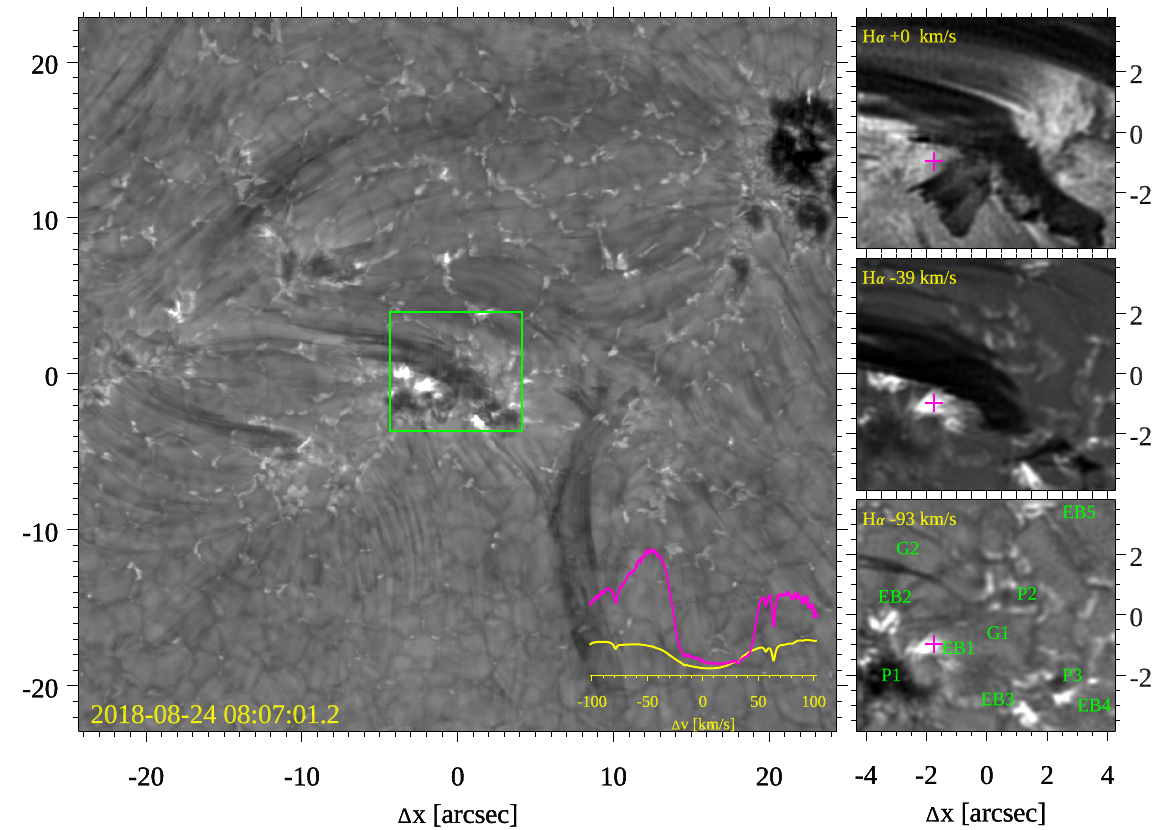}
\caption{
Overview of the MiHI observations of AR12720 on 24 Aug 2018. The left panel shows an image from the \Halpha\ wideband context imager, the green box outlines the MiHI field of view. The column at right shows MiHI images in the \Halpha\ core and two positions in the blue wing. Notable features are marked in the bottom right panel: micro-pores (P1, P2, and P3), clusters with EBs (EB1 -- 5), and elongated granules characteristic for strong flux emergence (G1 and G2). The pink cross marks the location in EB1 for which the spectral profile is shown in pink in the lower right corner of the large context image. The yellow profile is a reference spectrum averaged over the full MiHI FOV. An animation of this figure is available in the online material at \url{https://www.mn.uio.no/astro/english/people/aca/rouppe/movies/rouppe_mihi_fig02.mp4}.
}
\label{fig:overview}%
\end{figure*}

Figure~\ref{fig:overview} shows example MiHI images at three different spectral positions and the associated wideband context image. 
The wideband channel integrates over the full \Halpha\ pre-filter passband and shows a mix of both chromospheric and photospheric structures. 
A large system of chromospheric fibrils was crossing the MiHI FOV, the accompanying animation shows that they consist of long and narrow threads. 
Some chromospheric stuctures have strong Doppler shifts and are visible in the \Halpha\ wing images. 
The far blue wing MiHI image shows the clearest view at the underlying photosphere. 
Here we can identify three micro-pores, marked as P1, P2 and P3. 
Micro-pore P1 in the lower left corner is largest, P2 and P3 are smaller and less dark in the central area. They are surrounded by bright rims with bright points and have similar morphology as the small micro-pores and ribbons that were described by 
\citet{2004A&A...428..613B} 
and \citet{2005A&A...435..327R}. 

Throughout the 10 min duration of the observation, there are a number of EBs at different locations in the FOV. 
In the far blue wing MiHI image of Fig.~\ref{fig:overview}, five different sites with clusters of EBs are marked as EB1 -- 5. 
Cluster EB5 is not so bright in Fig.~\ref{fig:overview} but it can be seen with clear enhanced brightness earlier and later in the accompanying movie. 
The three MiHI panels are a nice illustration that EBs reside below the  chromosphere: there are no clear EB signatures in the line core image, while the far wing image shows an unobstructed view on the EB sites
\citep[this was earlier shown by][]{2011ApJ...736...71W}. 
In the intermediate wing position (middle panel), EBs are partly covered by overlying chromospheric structures. This is best seen in the movie. 

\begin{figure*}[!ht]
\centering
\includegraphics[width=16cm]{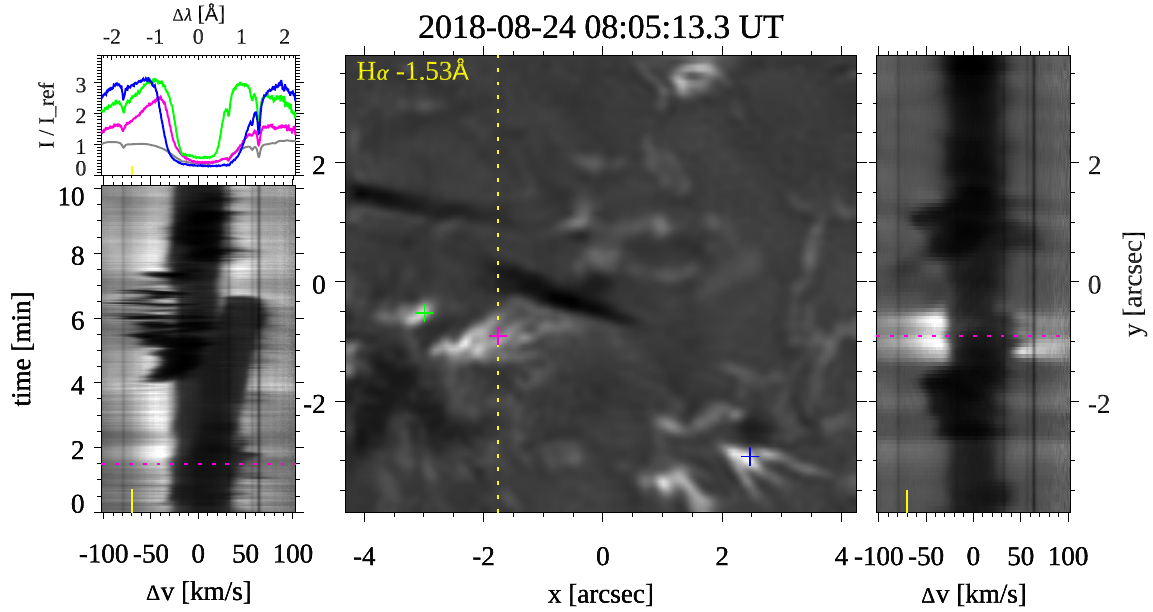}
\caption{
Example EB profiles. The middle panel shows a MiHI \Halpha\ blue wing image where colored crosses mark the locations for which spectral profiles are shown in the top left panel. Profiles are shown from EB clusters (see Fig.~\ref{fig:context}) EB1 (pink), EB2 (green) and EB4 (blue). The grey profile is averaged over the whole field of view and serves as reference. All profiles are normalised to the far blue wing intensity of the reference profile. The lower left panel shows the spectral evolution for the location marked with the pink cross as a $\lambda t$-diagram. The right panel shows the spectra along the dotted yellow line in the \Halpha\ wing image as a $\lambda y$-diagram. Horizontal pink dotted lines in the two diagrams mark the pink profile. Small vertical yellow markers indicate the spectral position of the \Halpha\ wing image. 
Two animations of this figure are available in the online material at \url{https://www.mn.uio.no/astro/english/people/aca/rouppe/movies/}: one showing the full \href{https://www.mn.uio.no/astro/english/people/aca/rouppe/movies/rouppe_mihi_fig03_tevolution.mp4}{temporal evolution} and one showing a \href{https://www.mn.uio.no/astro/english/people/aca/rouppe/movies/rouppe_mihi_fig03_linescan.mp4}{spectral line scan} stepping through the full spectral range.
}
\label{fig:ebexamples}%
\end{figure*}

Example EB profiles from three different EB clusters are shown in Fig.~\ref{fig:ebexamples}. 
They show the characteristic EB profile with a central absorption core, peak intensity in the wings at around $\pm$1~\AA\ offset from line center and decreasing intensity towards larger wavelength offset. 
These are very strong EBs as the peak intensity reaches up to 3 times or more the reference intensity from the average profile. This is more than double the typical threshold value that is used for automatic EB detection methods 
\citep[see][]{2019A&A...626A...4V}. 
The enhanced wing emission extends over a wide wavelength range, probably far beyond the $\pm$2.2~\AA\ wavelength range of the MiHI observations as one would infer from a naive visual extrapolation of the far wing profile trends. 
The animation of the spectral line scan that is accompanying Fig.~\ref{fig:ebexamples} is another vivid illustration that EBs are situated below the chromosphere as they become increasingly more covered by chromospheric structures for smaller offsets from line center. 
The other animation accompanying Fig.~\ref{fig:ebexamples} shows the temporal evolution at $-70$~\kms\ Doppler offset and detailed spectral evolution for a location in the EB1 cluster. 
The movie clearly shows the bursty character and rapid evolution of this EB cluster. 
A large number of small and faint blobs can be seen to be ejected. These blobs will be discussed in more detail further below. 

\subsection{Anomalously elongated granules}

\begin{figure*}[!ht]
\includegraphics[width=\textwidth]{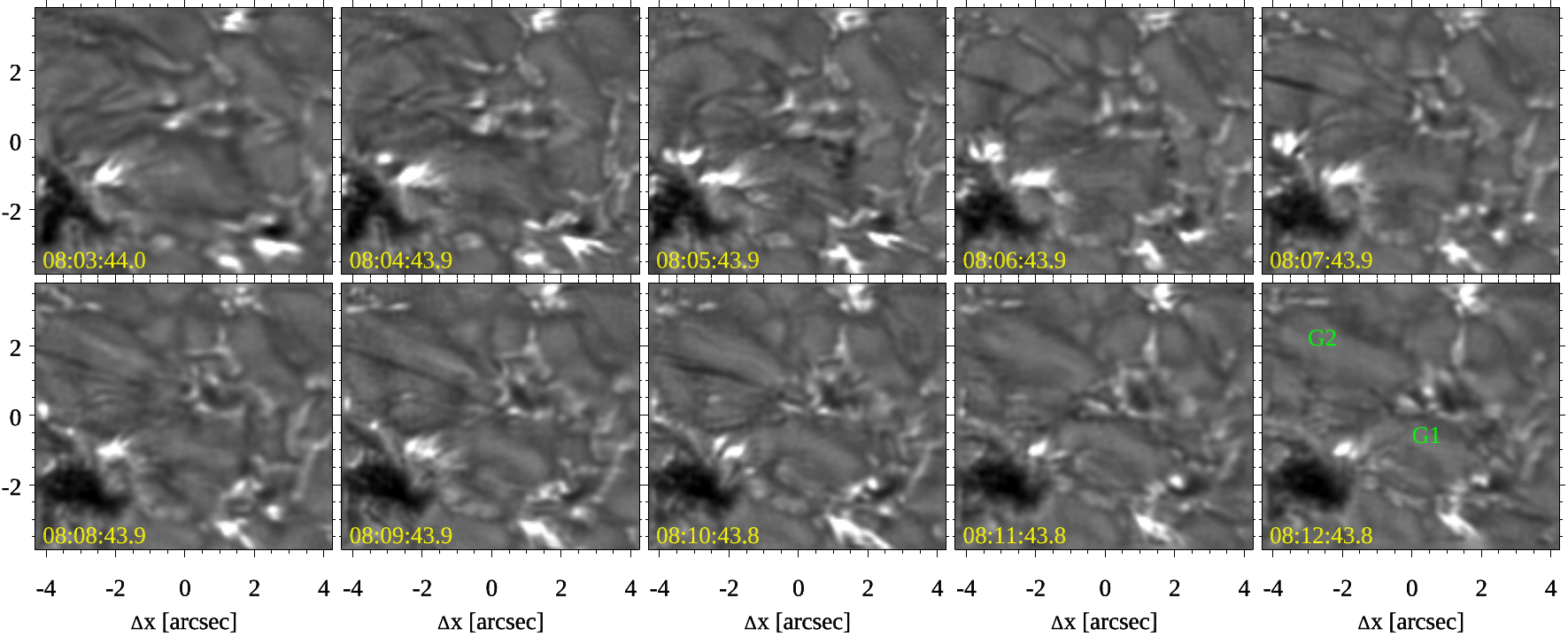}
\caption{
Evolution of anomalous elongated granules in a series of MiHI \Halpha\ far wing images. The images are summed over 10 spectral positions in both the blue and red wings. The labels G1 and G2 in the lower right panel mark the two prominent elongated granules (also see Fig.~\ref{fig:overview}).
An animation of this figure is available in the online material at \url{https://www.mn.uio.no/astro/english/people/aca/rouppe/movies/rouppe_mihi_fig04.mp4}.
}
\label{fig:hawing}%
\end{figure*}

\begin{figure*}[!ht]
\sidecaption
\includegraphics[width=12cm]{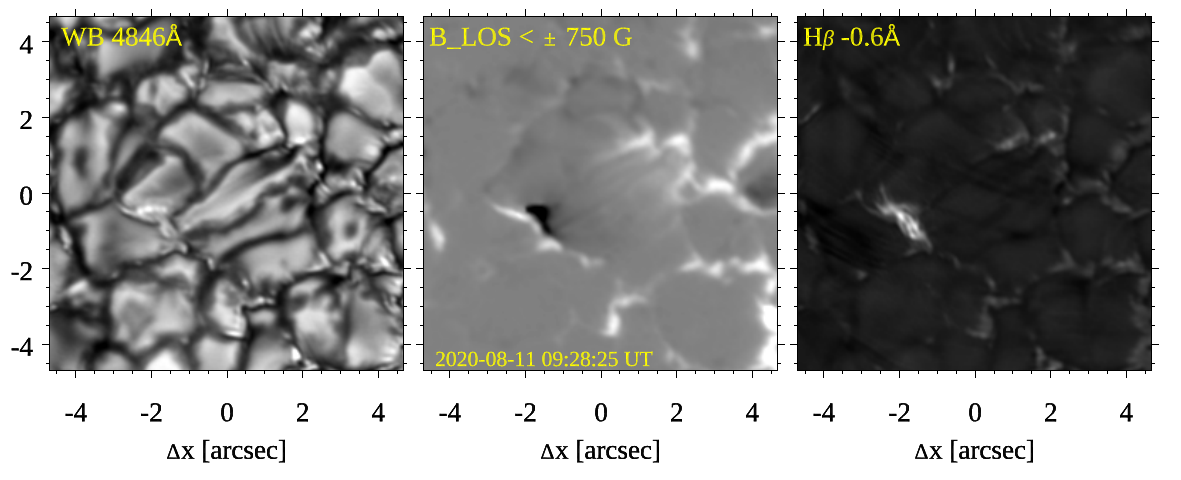}
\caption{
Episode of strong flux emergence with characteristic elongated granules in 11 Aug 2020 CHROMIS and CRISP observations. The left panel shows a wideband (FWHM 6.5~\AA) image centered on 4846~\AA. The middle panel shows a map of the magnetic field $B_\mathrm{LOS}$ derived from inversions of the \ion{Fe}{i}~6173~\AA\ line. The right panel shows an \Hbeta\ blue wing image at the wavelength offset of the peak of the typical EB profile. This image is scaled linearly between the minimum and maximum. 
An animation of this figure is available in the online material at \url{https://www.mn.uio.no/astro/english/people/aca/rouppe/movies/rouppe_mihi_fig05.mp4}.
}
\label{fig:textbook}%
\end{figure*}

The EBs in the EB1 and EB2 clusters are at the ends of what appear to be abnormally long and elongated granules, marked as G1 and G2 in the far wing panel of Fig.~\ref{fig:overview}. 
The evolution of these elongated granules is shown in summed \Halpha\ far wing images in Fig.~\ref{fig:hawing} and associated animation.
Anomalously elongated granules can be regarded as a characteristic sign of emergence of strong magnetic flux
\citep{2008ApJ...687.1373C, 
2010ApJ...724.1083G, 
2014ApJ...781..126O, 
2014LRSP...11....3C}. 
The small and crowded FOV of MiHI and the relatively low granulation contrast of the \Halpha\ pre-filter, make it not easy to recognize the elongated granules. 
As reference, Fig.~\ref{fig:textbook} shows CRISP $B_\mathrm{LOS}$ magnetograms and CHROMIS observations of a similar episode of strong magnetic flux emergence in an active region. 
The accompanying movie shows the emergence of magnetic flux with two opposite polarity patches moving apart. 
These two opposite polarity patches reside at the end points of an anomalously elongated granule. 
The WB 4846~\AA\ image in Fig.~\ref{fig:textbook} shows that this granule is divided in the middle along its major axis by a darker lane.
\citet{2014LRSP...11....3C} remarked that flux emergence simulations show that such dark lanes are coincident with upflows. 
The \Hbeta\ wing image shows clear EBs when the negative polarity patch collides with the pre-existing positive polarity patches.
These EBs show similar rapid and bursty evolution as in the MiHI data but at much slower temporal sampling (17~s). 
Some of the time steps show the ejection of small faint blobs.

\subsection{Plasmoid-like blobs}

\begin{figure*}[!ht]
\centering
\includegraphics[width=8.95cm]{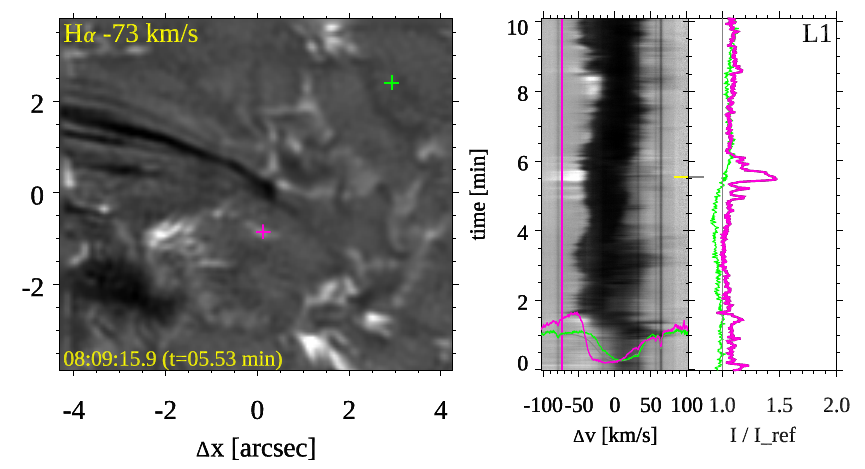}
\includegraphics[width=8.95cm]{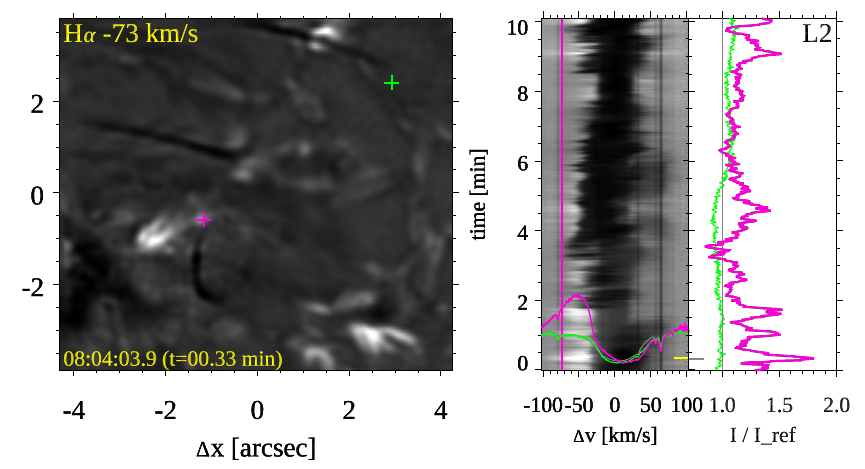} \\
\includegraphics[width=8.95cm]{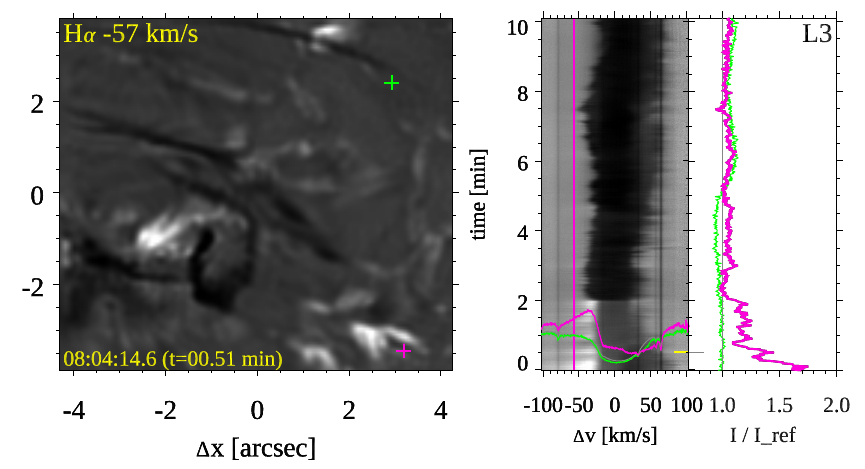}
\includegraphics[width=8.95cm]{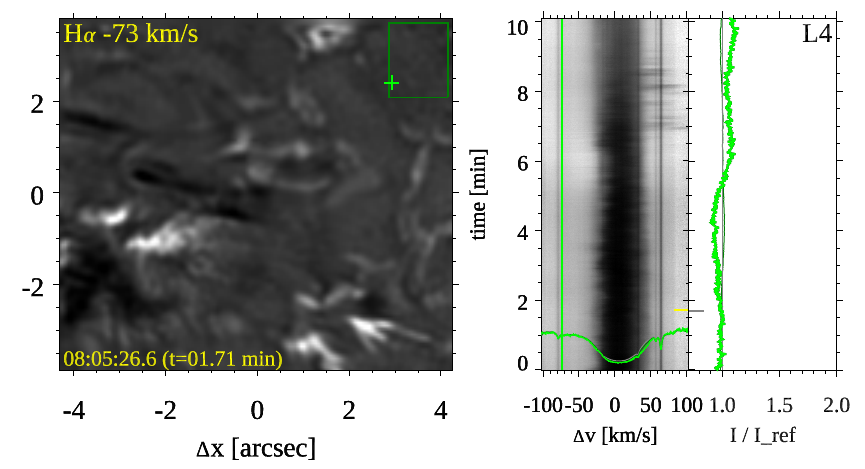} \\
\caption{
Examples of rapid temporal variability in light curves close to EB clusters. The top row shows two examples in the vicinity of cluster EB1, the bottom left shows an example close to EB4. In the left \Halpha\ wing images, a pink cross marks the location for which the $\lambda t$-diagram and the light curve (pink) are shown at right. The \Halpha\ wing image is shown for a time when a blob can be seen at this location. Line profiles from this time are overplotted on the $\lambda t$-diagram. The light curve is at the wavelength marked with the pink vertical line in the $\lambda t$-diagram and noted in the top left of the MiHI wing image. The wavelength is selected to avoid excursions of the \Halpha\ line core as much as possible so that the light curve variations can be mostly attributed to EB variability. The green light curve serves as reference and is drawn from the location marked with the green cross. The light curves are normalized by a reference light curve that is constructed by spatial averaging over the rectangular area in the top right corner of the FOV that is marked in dark green in the bottom right. This spatial average light curve is shown in dark green in the bottom right light curve panel (L4) and is normalized to its temporally averaged value. A background grey line serves as a guide to $I/I_\mathrm{ref}=1.0$.
Animations of the figures for \href{https://www.mn.uio.no/astro/english/people/aca/rouppe/movies/rouppe_mihi_fig06_l1.mp4}{L1}, 
\href{https://www.mn.uio.no/astro/english/people/aca/rouppe/movies/rouppe_mihi_fig06_l2.mp4}{L2}, and 
\href{https://www.mn.uio.no/astro/english/people/aca/rouppe/movies/rouppe_mihi_fig06_l3.mp4}{L3} are available in the online material at \url{https://www.mn.uio.no/astro/english/people/aca/rouppe/movies/}.
}
\label{fig:lightcurves}%
\end{figure*}

Figure~\ref{fig:lightcurves} shows light curves from locations where bright blobs can be seen crossing. 
These locations are at quite a distance from the EB cluster from where the blobs appear to be originating. 
For the light curve in the upper left (marked L1), at time $t=5.53$~min, an elongated blob can be seen to pass under the location marked with the pink cross in the \Halpha\ wing image. Its FWHM size along the longest dimension is 0\farcs33 and along the shortest dimension 0\farcs19. As can be seen in the accompanying movie, it appears to be connected to the EB1 cluster which has its root at about 2\arcsec\ distance. For a period of about 17~s, the light curve at $-73$~\kms\ Doppler offset shows clear enhancement, peaking at about 1.5 times the reference intensity. The spectral profile (shown as pink profile in the $\lambda t$-diagram) shows clear EB-like wing enhancement.  

The light curve in the upper right (L2) is from a location closer to EB1 and shows more bursty behaviour with short episodes of enhancements than L1. 
The accompanying movie shows that these enhancements come from many small blobs passing under the pink cross. 
At $t=0.33$~min, the blob that is crossing has FWHM sizes 0\farcs26 by 0\farcs18. The line profile shows a peak EB wing enhancement of more than a factor 2.  

The light curve L3 close to EB4 near the edge of the FOV shows bursty intensity peaks only in the first 2~min. 
The blob that passed at $t=0.51$~min was round with a FWHM of 0\farcs21 and made the \Halpha\ $-57$~\kms\ light curve peak for about 12~s. 
The blob has an EB-like spectral profile with peak intensity larger than 1.5. 
The EB activity in the EB4 cluster continued for a few more minutes and the light curve remains slightly enhanced until about $t=5$~min. For the last 3~min all EB activity is gone and the light curve is very similar as the quiet reference light curve. 
The L4 light curve shows the reference light curve that is shown in the other panels in comparison with a light curve averaged over an area of 1\farcs30$\times$1\farcs63. 

Two additional light curves are shown in Fig.~\ref{fig:more_lightcurves}. 
Light curve L5 is close to EB1 and bright blobs pass for about 50~s. At $t=1.02$~min, the passing blob has shortest FWHM 0\farcs12, and longest 0\farcs25. 
Light curve L6 is close to EB5 near the edge of the FOV. 
%

\begin{figure*}[!ht]
\sidecaption
\includegraphics[width=12cm]{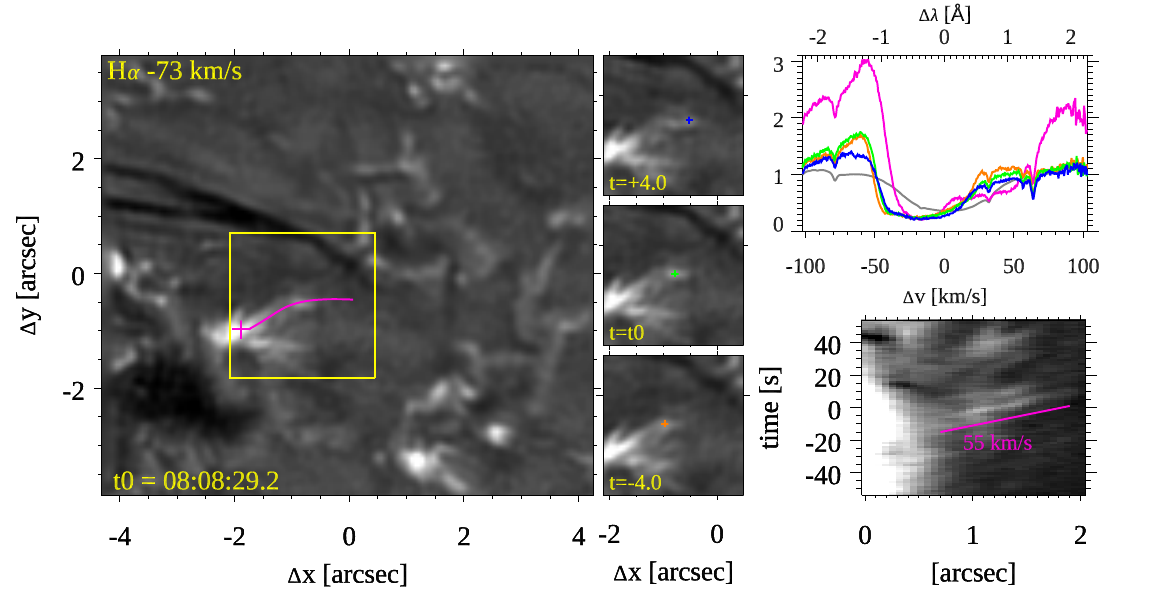}
\caption{
Apparent motion of plasmoid-like blobs.
The pink line in the large \Halpha\ wing image at left traces the trajectory of some bright blobs. The $s t$-diagram for this trajectory is shown in the bottom right. The column of smaller images shows three different time steps where colored crosses mark the moving blob for which the respective spectral profiles are shown in the top right panel. This blob has an apparent speed of about 55~\kms, a speed that is indicated by a straight pink line in the $s t$-diagram. 
The pink profile in the top right panel is a reference EB profile that is marked with the large pink cross in the large image at left. The cross is near the origin of the trajectory for the $s t$-diagram. The grey profile is a reference profile that was constructed by averaging over the full field of view.  
}\label{fig:spacetime}%
\end{figure*}

\begin{figure*}[!ht]
\sidecaption
\includegraphics[width=12cm]{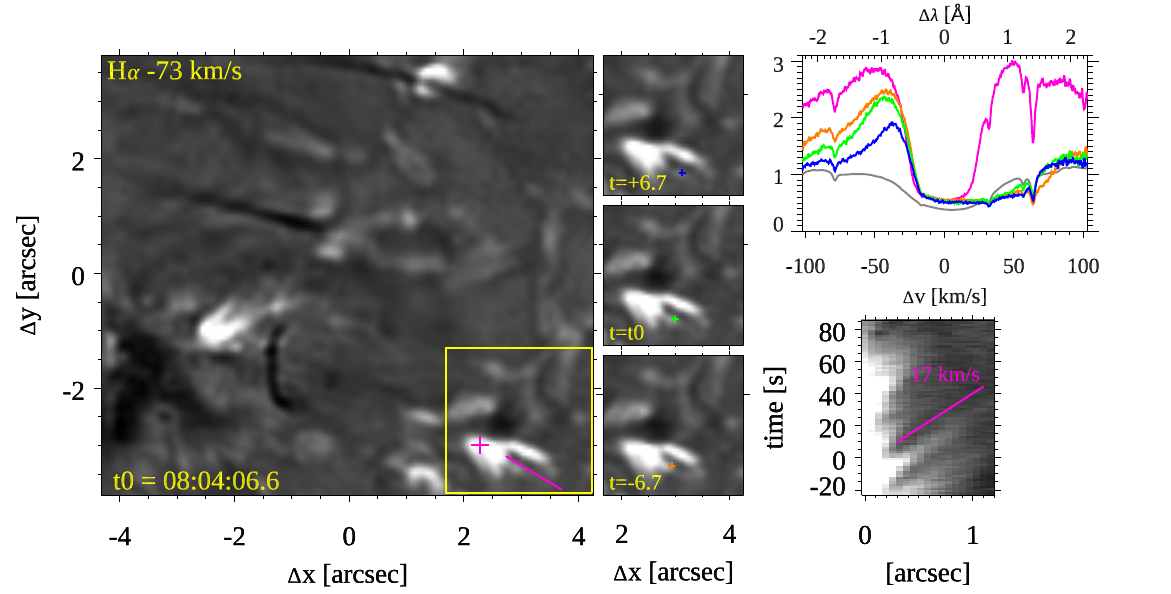}
\caption{
Another example of the apparent motion of plasmoid-like blobs.
Same format at Fig.~\ref{fig:spacetime}.
}\label{fig:spacetime2}%
\end{figure*}

\begin{figure*}[!ht]
\sidecaption
\includegraphics[width=12cm]{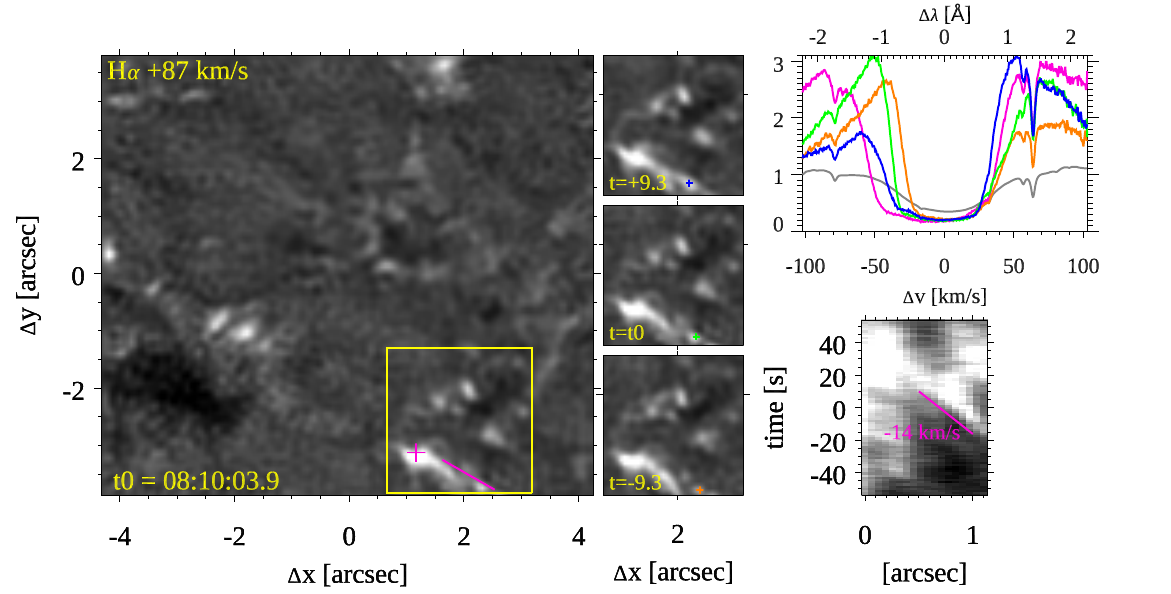}
\caption{
Another example of the apparent motion of plasmoid-like blobs.
Same format at Fig.~\ref{fig:spacetime}.
}\label{fig:spacetime3}%
\end{figure*}

The blobs are small, faint, and dynamic. We managed to make an estimate of the apparent speed for a few blobs by tracing their trajectories and constructing space-time diagrams.
In Fig.~\ref{fig:spacetime} the curved trajectory of an elongated blob (0\farcs28 by 0\farcs18) is shown in the large \Halpha\ wing image. 
The $s t$-diagram in the lower right shows the motion of the blobs as a bright inclined streak.
The inclination of this streak is equivalent to an apparent speed of about 55~\kms. 
The $s t$-diagram shows more inclined streaks caused by other blobs passing the trajectory at other times. Their inclination is similar but it should be noted that this particular trajectory may not optimally trace the path of other blobs than the one centered at $t=0$. 
The column of small \Halpha\ wing images shows the position of the blob at three instances of time. 
The blob's spectral profiles are shown in the profile panel. The profiles are EB-like and have peak wing enhancement at about 1.7 times the reference intensity. 

The blob that is traced in Fig.~\ref{fig:spacetime2} is associated with EB4 and is moving at an apparent speed of about 17~\kms. 
The profiles of this blob have stronger peak intensity than the blob in Fig.~\ref{fig:spacetime} with a maximum well above 2. 
These two blobs appear to be moving away from their respective EB sites. 
The blob traced in Fig.~\ref{fig:spacetime3} is associated with EB3 and appears to be moving towards the EB site with an apparent speed of 14~\kms. The profiles of this blob are very strong EB profiles. 
The $s t$-diagram in Fig.~\ref{fig:spacetime4} traces very fast blobs that appear to move at a speed of about 77~\kms. 

Ellerman bombs occur in the deep atmosphere below the chromospheric canopy and it seems unlikely that one can observe a complete \Halpha\ profile from the EB site itself that is not affected by opacity from unrelated plasma in the overlying chromospheric fibrils. 
A ``naked'' EB profile from a site without overlying chromosphere would presumably have a single emission peak without the strong central absorption around nominal \Halpha\ line center that effectively produces the characteristic pair of peaks in the canonical EB profile. 
Flares can produce pure emission lines in \Halpha\ and the peak wavelength position is a rather straightforward Doppler diagnostic in flare ribbons. 
The position of the EB peaks in profiles like in the examples shown in Fig.~\ref{fig:ebexamples} are meaningless as a Doppler diagnostic and are clearly affected by the Doppler shift and width of the overlying chromospheric structures. 
However, for a few blobs, we find highly asymmetric profiles with a single emission peak in one of the wings and a seemingly unaffected wing shape at the opposite side of the line core.

\begin{figure*}[!ht]
\centering
\includegraphics[width=7.9cm]{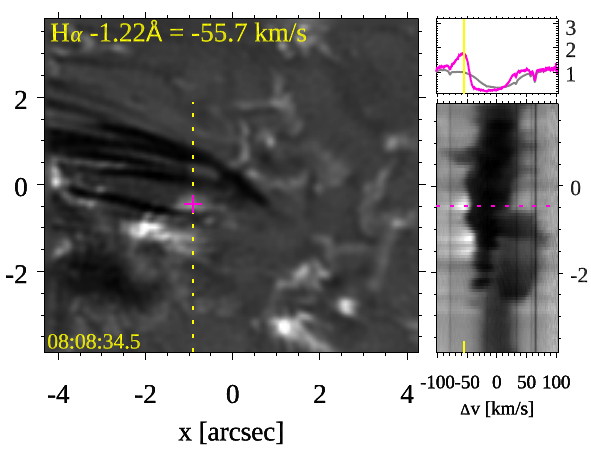}
\includegraphics[width=7.9cm]{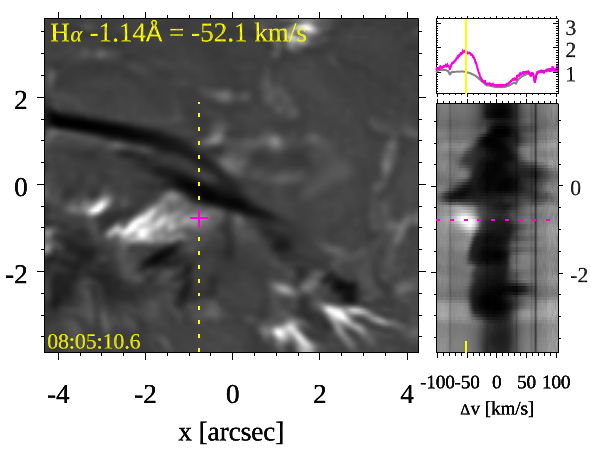} \\
\includegraphics[width=7.9cm]{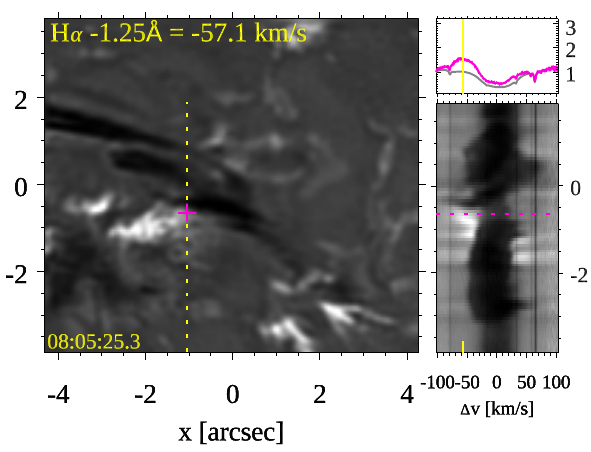}
\includegraphics[width=7.9cm]{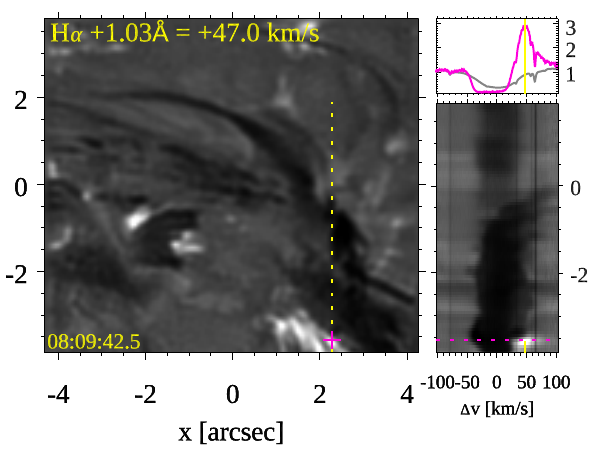} \\
\caption{
Four examples of highly asymmetric spectral profiles that are suggestive of Doppler shifted emission peaks. The pink cross in the left \Halpha\ wing images marks the location for which the full spectral profile is shown in the top right (pink profile). The grey profile is a reference profile average over the full MiHI field of view. The spectral profiles are normalised to the intensity at the shortest wavelength of the reference profile. The vertical yellow dashed line marks the location for which the $\lambda y$-diagram is shown at right. 
}\label{fig:dopplershift}%
\end{figure*}

Figure~\ref{fig:dopplershift} shows four examples. 
The top left example has a spectral profile with an emission peak with $I/I_\mathrm{ref} > 1.7$ at a Doppler offset of about $-$56~\kms. 
The red wing does not show significantly enhanced emission. The \Halpha\ line core absorption seems to be blue shifted with the line minimum at about $-20$~\kms. The line core absorption profile is slightly wider than the rather quiet profiles near the bottom of the $\lambda y$-diagram below $y=-3\arcsec$, but the red wing of the blob profile is certainly not as much affected by absorption as the wide \Halpha\ profiles for $y=-0\farcs6$ to $-$2\farcs5. The red wing profiles in this part of the $\lambda y$-diagram are totally dominated by the downflow of a surge-like event. 
The spectral profile for this example seems to not have a red wing EB peak and it is therefore suggestive that the EB emission comes from strongly blue shifted plasma. 
The $-56$~\kms\ offset of the peak intensity is probably an upper limit since it seems likely that the position of the emission peak is affected by the blue shifted \Halpha\ absorption core. 

The upper right example is from a different time in the same region and shows peak intensity at about $-52$~\kms. The line core absorption has similar width as the reference profile and does not appear to be Doppler shifted. The red wing shows no sign of EB emission and closely follows the reference profile. 
The red wing of the lower left example is also not affected and this profile has a rather weak line core absorption that has higher intensity than the reference profile. The peak intensity is at about $-57$~\kms. 

The fourth example in the lower right is from a different region and is associated with the EB3 site where blobs can be seen moving towards the EB root site (see Fig.~\ref{fig:spacetime3}).
The spectral profile shows a very strong emission peak of nearly 3 at a Doppler offset of +47~\kms.
There is very strong \Halpha\ line core absorption which is blue shifted with line minimum at about $-30$~\kms. 
This can be attributed to a chromospheric upflow above the blob and affects the inner wing only. 
The far blue wing however does not show any enhanced emission. 
The red shift of this emission peak is suggestive of a downflow which is consistent with the apparent inflow that can be seen in this region. 
The blue shifts of the emission peaks of the other three examples indicate outflows that are consistent with the apparent motion of blobs away from the EB1 site. 

\section{Discussion and conclusion}
\label{sec:discussion}

We analyzed ultra-high resolution \Halpha\ observations from a prototype of the MiHI microlens based integral field spectrograph targeting the heart of a young emerging active region. 
The dataset combines unprecedented simultaneous high resolution in three domains: spatial resolution better than 0\farcs2, spectral resolution of $R\approx 315\,000$ (about 20~mÅ at 6563~\AA), and a temporal cadence of 1.33~s.  
Context SDO and IRIS observations showed that there was a high level of magnetic flux emergence and activity related to interactions between newly emerged and pre-existing ambient flux. 
At the photospheric level, the MiHI observations show the evolution of at least two anomalously elongated granules that are characteristic of strong magnetic flux emergence. 
A high level of EB activity was found to be associated with these elongated granules and in total five sites with more or less continuous and strong EB activity were found over the 10~min time series and 8\farcs6~$\times$~7\farcs7 MiHI FOV. 
These EB sites were clearly situated below the chromosphere as they were mostly visible in the far \Halpha\ wings and completely covered in spectral images around the line core. 
The EB activity displayed high variability on a timescale of seconds, both in terms of wing intensity and in terms of spatial morphology and extend of the area covered in the FOV. 
The EB areas had clear substructure and there were multiple episodes where small blobs could be observed to detach from the EB sites and move away at high speed. 
These blobs were found to have sizes below 0\farcs5 and down to the diffraction limit, we measured FWHM sizes between 0\farcs1 -- 0\farcs4. 
From $s t$-diagrams, the apparent speeds of 4 blobs were measured, and we found approximate speeds between 14 and 77~\kms. 
The blobs had EB-like spectral profiles with peak wing enhancement often well above 1.5 times the reference wing intensity, surpassing common enhancement thresholds that are used for automatic detection of EBs 
\citep[][]{2019A&A...626A...4V}. 
We found a few examples where the blob spectral profiles have only one emission peak and the opposite spectral wing appears to be unaffected by overlying chromospheric absorption. 
For these examples, the Doppler offset of the peak intensity could serve as velocity diagnostic of the EB emitting plasma. 
The Doppler offsets range between 47 and 57~\kms. These values should be regarded as upper limits since the peak profile and position is most likely affected by the overlying chromosphere. Still, these single peak profiles are a cleaner Doppler diagnostic than the canonical double peaked EB profiles. 
These Doppler velocities have values that are consistent with the apparent speeds. Moreover, we find blue shifts for blobs that appear to move away from their EB root sites and a red shift for a blob that is part of a group of blobs that is moving towards the EB site. 

High temporal variability in EBs was earlier found in 1~s cadence wideband \Halpha\ observations 
\citep{2011ApJ...736...71W, 
2016A&A...592A.100R}. 
Since these observations were filtergram data, they lacked the extensive spectral coverage of the MiHI data we analyzed here. 
The active region EBs analyzed in 
\citet{2011ApJ...736...71W} 
were found to have substructure features moving up with apparent speeds between 11 and 60~\kms, and down between 7 and 18~\kms. 
Our measurements for apparent speed and inferred Doppler velocities are consistent with these values. 

\citet{2017ApJ...851L...6R} 
observed fine structure and rapid variability in CHROMIS \ion{Ca}{ii}~K observations of EBs associated with magnetic flux emergence. 
Bright blobs with FWHM sizes smaller than 0\farcs15 were found in spectral images at Doppler offsets of about 40~\kms. 
The temporal cadence was longer than 10~s and no reliable measurements of the apparent velocity could be made. 
The EBs were associated with UV bursts and the IRIS \ion{Si}{iv} profiles that were cospatial and cotemporal with the \ion{Ca}{ii}~K blobs were of non-Gaussian and triangular shape. 
These types of profiles have been linked to magnetic reconnection driven by the plasmoid instability
\citep{2015ApJ...813...86I}. 
%
\citet{2020ApJ...901..148G} 
simulate the onset of fast reconnection in a thinning current sheet that breaks up in multiple plasmoids. As the plasmoids grow and coalesce, they adopt a broad range of velocities that collectively produce wide and non-Gaussian \ion{Si}{iv} profiles. The dynamic evolution of the simulated \ion{Si}{iv} profiles matched well with high cadence IRIS observations of UV bursts. 
\citet{2017ApJ...851L...6R} 
compared the \ion{Ca}{ii}~K and \ion{Si}{iv} observations with synthetic observables computed from a 2D Bifrost simulation of magnetic flux emergence
\citep{2017ApJ...850..153N}. 
In this simulation, plasmoids occur in the current sheet that forms during the coalescence of the newly emerging field with the pre-existing canopy.
The favorable similarity between the synthetic and observed spectral line profiles supports the idea that the small-scale and highly dynamic blobs in the SST observations are in fact plasmoids that are produced in reconnection in the deep solar atmosphere. 
A similar SST observation was reported by
\citet{2021A&A...647A.188D} 
who tracked a number of $\sim$0\farcs2 sized blobs in the blue wing of the \ion{Ca}{ii}~K line being ejected from a region with strong reconnection, also associated with magnetic flux emergence. The blobs were identified at a Doppler offset of $-100$~\kms\ and their apparent speed in the $\sim$9~s cadence time sequence was estimated to range between 70 and 100~\kms.

The similarities with these earlier high-resolution observations support the interpretation that the blobs in the \Halpha\ MiHI observations are plasmoids. 
A strong advantage of MiHI is that it removes the classic limitations of either tunable filtergraphs like CHROMIS or scanning spectrographs like IRIS and delivers simultaneous high resolution in many domains. 
These \Halpha\ observations illustrate that this allows to better characterize the fundamental process of magnetic reconnection in the deep solar atmosphere.
While the interpretation of the \Halpha\ line is challenging due to its complex spectral line formation, it is possible to get a clearer view on EB sites in the shorter wavelength Balmer lines
\citep{2020A&A...641L...5J}. 
Not only does the shorter wavelength allow for higher spatial resolution and higher intensity contrast, the higher order Balmer lines are also less affected by opacity in the overlying chromosphere. 
The CHROMIS observations of the EB in Fig.~\ref{fig:textbook} show that plasmoid-like blobs can also be detected in \Hbeta.
It is an exciting prospect that the next generation of 4-m class telescopes should be capable of zooming in on even smaller spatial scales.
Among many other elementary questions, this may resolve the question whether these $\sim$0\farcs2 blobs really are single plasmoids or rather conglomerates of multiple, smaller plasmoids. 
Our observations make a strong case that a MiHI like integral field spectrograph at a 4-m class telescope is a powerfull combination, that, aided with modern non-LTE inversions 
\citep[e.g.,][]{2019A&A...627A.101V} 
and advanced numerical simulations
\citep[e.g.,][]{2022ApJ...935L..21N}, 
is ideally suited to make fundamental progress in the understanding of the physics of the solar atmosphere. 

\begin{acknowledgements}
The Swedish 1-m Solar Telescope is operated on the island of La Palma
by the Institute for Solar Physics of Stockholm University in the
Spanish Observatorio del Roque de los Muchachos of the Instituto de
Astrof{\'\i}sica de Canarias.
The Institute for Solar Physics is supported by a grant for research infrastructures of national importance from the Swedish Research Council (registration number 2021-00169).
IRIS is a NASA small explorer mission developed and operated by LMSAL with mission operations executed at NASA Ames Research center and major contributions to downlink communications funded by ESA and the Norwegian Space Centre.
This research is supported by the Research Council of Norway, project numbers 250810, 
325491, 
and through its Centres of Excellence scheme, project number 262622.
This project was supported by the European Commission's FP7 Capacities Program under the Grant Agreements No. 212482 and No. 312495.
It was also supported by the European Union's Horizon 2020 research and innovation program under the Grant Agreements No. 653982 and No. 824135. 
This project has received funding from the European Research Council (ERC) under the European Union's Horizon 2020 research and innovation program: grant agreement No 695075 (MvN) and SUNMAG, grant agreement 759548 (JdlCR).
This work benefited from discussions during the workshop “Studying magnetic-field-regulated heating in the solar chromosphere” (team 399) at the International Space Science Institute (ISSI) in Switzerland. 
We thank Dr. D.~N{\'o}brega~Siverio for illuminating discussions.
We made much use of NASA's Astrophysics Data System Bibliographic Services.
\end{acknowledgements}


\begin{appendix}

\section{More examples}
\label{app:more_examples}

\begin{figure*}[!hb]
\centering
\includegraphics[width=8.95cm]{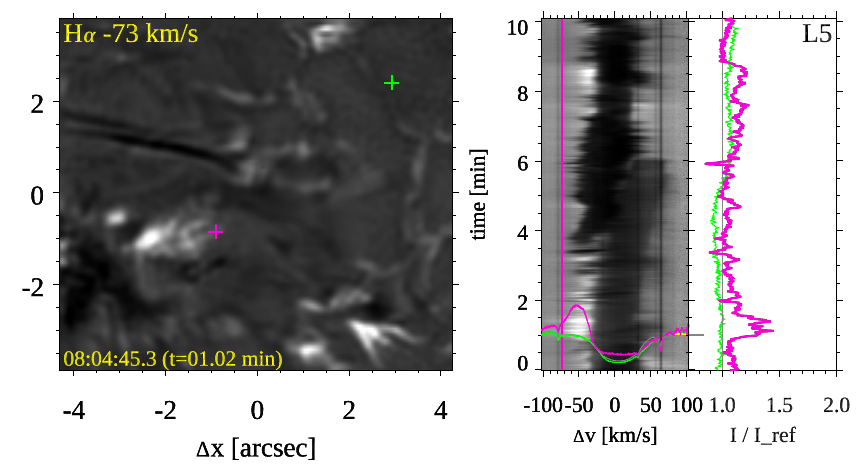}
\includegraphics[width=8.95cm]{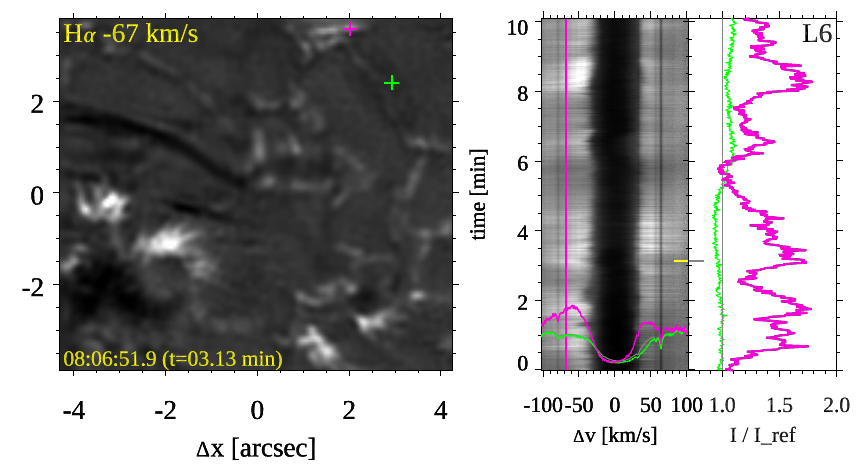} \\
\caption{
Two additional light curves, see Fig.~\ref{fig:lightcurves}. Light curve L5 is close to EB cluster EB1, L6 close to EB5 (see Fig.~\ref{fig:overview}. 
Animations of this figure are available in the online material at \url{https://www.mn.uio.no/astro/english/people/aca/rouppe/movies/}: 
\href{https://www.mn.uio.no/astro/english/people/aca/rouppe/movies/rouppe_mihi_figa1_l5.mp4}{L5}, and 
\href{https://www.mn.uio.no/astro/english/people/aca/rouppe/movies/rouppe_mihi_figa1_l6.mp4}{L6}.
}
\label{fig:more_lightcurves}%
\end{figure*}

\begin{figure*}[!hb]
\sidecaption
\includegraphics[width=12cm]{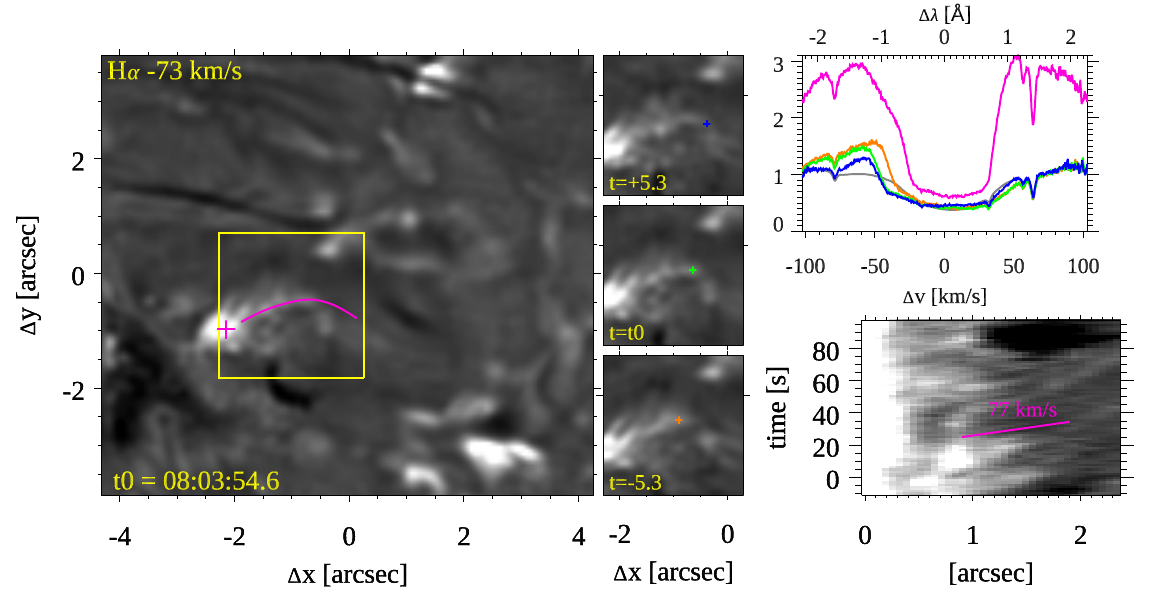}
\caption{
Another example of the apparent motion of plasmoid-like blobs.
Same format at Fig.~\ref{fig:spacetime}.
}\label{fig:spacetime4}%
\end{figure*}

Two more lightcurve are shown in Fig.~\ref{fig:more_lightcurves}, one close to EB cluster EB1 and one close to EB5. 
One additional example of the apparent motion of plasmoid-like blobs is shown in Fig.~\ref{fig:spacetime4}.

\end{appendix}

\end{document}